\documentclass[amssymb,amsmath,twocolumn,preprintnumbers,prb]{revtex4}
\usepackage{graphicx}
\usepackage{dcolumn}
\usepackage{textcomp}

\begin{document}

\newcommand {\ket}[1]{\ensuremath{| #1 \rangle}}
\newcommand {\bra}[1]{\ensuremath{\langle #1 |}}

\title{Analysis and Geometric Optimization of Single Electron Transistors for Read-Out in Solid-State Quantum Computing.}
\date{\today}
\author{Vincent I. Conrad\footnote{Author to whom correspondence should be addressed.}, Andrew D. Greentree, David N. Jamieson, Lloyd C.L. Hollenberg}
\affiliation{Centre for Quantum Computer Technology, School of
Physics, University of Melbourne, Vic. 3010, Australia}

\begin{abstract}
\normalsize

The single electron transistor (SET) offers unparalled opportunities as a nano-scale electrometer, capable of measuring sub-electron charge variations.  SETs have been proposed for read-out schema in solid-state quantum computing where quantum information processing outcomes depend on the location of a single electron on nearby quantum dots.  In this paper we investigate various geometries of a SET in order to maximize the device's sensitivity to charge transfer between quantum dots.  Through the use of finite element modeling we model the materials and geometries of an Al/Al$_2$O$_3$ SET measuring the state of quantum dots in the Si substrate beneath.  The investigation is motivated by the quest to build a scalable quantum computer, though the methodology used is primarily that of circuit theory.   As such we provide useful techniques for any electronic device operating at the classical/quantum interface.  

\end{abstract}

\maketitle

\section{Introduction}
The single electron transistor (SET)\cite{Fulton_1987,Devoret_2000} has presented itself as a very interesting device in the field of classical computing.  Systems using SETs as logic devices have already been proposed\cite{Uchida_2003}.  SETs also present the possibility of utilizing the spin orientation of electrons in the currents passing through them.  This leads to their use in the field of spintronics\cite{Smet_2002,Prinz_1998}, opening a plethora of possibilities in which this extra degree of freedom can be used to encode information.  A SET can also be used as a highly sensitive electrometer capable of measuring charge variations as small as $10^{-6}{q_e}/\sqrt{\rm {Hz}}$ which is close to the quantum limit\cite{Devoret_2000}.  Our research is motivated by the use of the SET to perform solid-state quantum computer read-out, though the techniques employed here are applicable to any application using a SET as an electrometer. 

The field of quantum computing began in the 1980s when researchers began to theoretically investigate how to apply quantum mechanics to information processing.  The first application envisioned for such quantum information processing was as a physics simulator, presented by Feynman in 1982 when he noted that classical systems can not simulate quantum systems efficiently\cite{Feynman_1982}.  In 1985 Deutsch developed a quantum algorithm\cite{Deutsch_1985} demonstrating the potential for quantum information processing in its own right.  However it was not until the discovery by Shor in 1994 of a quantum algorithm for polynomial-time factorization of prime numbers\cite{Shor_1994} that the possibilities for quantum information processing other than simulating physics began to capture the mainstream imagination.  Shor's work provided much impetus to the field due to the ramifications to RSA encryption.  Grover developed another quantum algorithm\cite{Grover_1997} in 1997 for improved database searching.  This has spurred further interest in the development of architectures for quantum computers, and other useful quantum algorithms.  Since that time the race to build a working quantum computer has taken off across the globe.  

The essential difference between classical computers and quantum computers is the replacement of the dualistic on/off bit (0 or 1), with the quantum-bit (qubit) which is capable of being in both the on and off state simultaneously ($\alpha$\ket{0}+$\beta$\ket{1} with $|\alpha |^{2}+|\beta |^2 = 1$ and $\alpha,\beta \in \mathbb{C}$).  Qubits may also make use of the purely quantum mechanical property of entanglement, which is the source of the unique computational power of quantum information processing\cite{Blume-Kohout_2002}.    The operation of any quantum computer requires that qubits can be individually `rotated' (i.e. the values of $\alpha$ and $\beta$ can be manipulated for each qubit) and that two qubit interactions can be controlled (to allow for multi-qubit logical operations such as CNOT).  While these are the essential qubit requirements for universal information processing, an actual physical implementation requires further considerations\cite{DiVincenzo_1995}.

One of the biggest difficulties in quantum information processing is that the encoded information is extremely fragile.  A delicate balance between optimizing the desired entanglement between qubits, and minimizing the undesired entanglement between the qubit and its environment must be achieved.  Any interaction a qubit might have with its environment causes the information stored on the qubit to ``leak'' out.  This is referred to as decoherence.  Until the development of quantum error correction algorithms\cite{Steane_1996,Shor_1996} it was believed by many that this decoherence would prevent the construction of any practical quantum computer.  Even though constructing a quantum computer is still at the limits of current fabrication technologies, these error correction algorithms show that a working quantum computer is (in principle) possible, as long as the error rate on the qubits can be kept below some threshold (a commonly used estimate\cite{Preskill_1997} is ~$10^{-4}$). 

\begin{figure}[tb!]
\centering
\includegraphics[width=7cm]{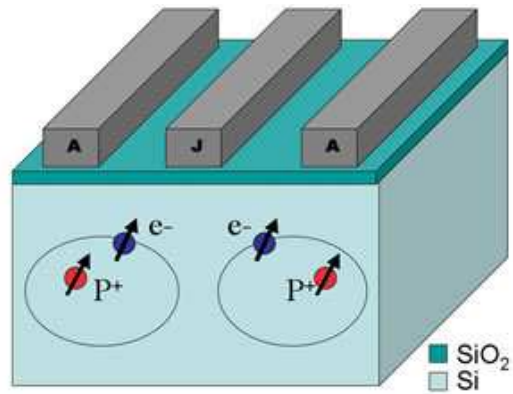}
\caption{Schematic of the Kane quantum computer.  The qubit states are encoded in the orientation of $^{31}$P nuclear spin.  Quantum information is mediated by the donor valence electrons.  A gates control single qubit rotations via the hyperfine coupling between a valence electron and its nucleus.  J gates control qubit interaction.  Read-out is via a SET (not shown) in the vicinity of a P donor in a known state.}
\label{fg:Kane_qc}
\noindent 
\end{figure}

\begin{figure}[tb!]
\centering
\includegraphics[width=7cm]{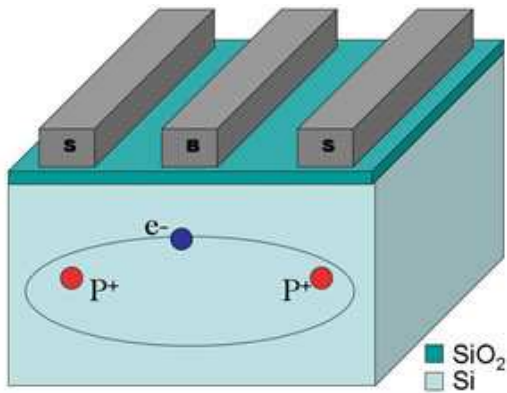}
\caption{Schematic of the charge-qubit quantum computer.  The qubit states are encoded in the location of a shared P-donor valence electron.  S gates control the symmetry of potential wells about P$^+$ donors.  The B gate creates a barrier potential to control tunneling between the donor pair.  Read-out is via a SET (not shown) used to determine the location of the shared electron on the P-P$^+$ donor pair. }
\label{fg:charge_qc}
\noindent 
\end{figure}

In this paper we are concerned with the read-out stage of a silicon based quantum computers.  Read-out constitutes projecting the quantum information stored in a qubit onto the sub-space of classical information of the macroscopic measuring device.  Qubit read-out will, in general, be probabilistic.  The read-out process for the state $\alpha$\ket{0}$+\beta$\ket{1}  will determine a 0 or 1 with respective probabilities of $|\alpha |^2$ and $|\beta |^2$.  We consider the read-out process for two proposed Si:P solid state quantum computers (SSQC), the Kane SSQC\cite{Kane_1998} (Fig.\ref{fg:Kane_qc}) and the charge-qubit SSQC\cite{Hollenberg_2004}  (Fig.\ref{fg:charge_qc}).  Much theoretical work has been done on optimizing the read-out process based on how to encode the quantum information and which degree of freedom of the qubit we choose to measure\cite{Vrijen_2000,Hill_04,de_Soussa_2004}.  Most of these proposals essentially require the determination of the location of a single electron for read-out.  We have focused on the SET design in order to optimize it as an electrometer for determining this electron location.

The Kane quantum computer (Fig.\ref{fg:Kane_qc}) uses the nuclear spin orientation (with respect to an external global magnetic field) of a single P dopant in silicon as its qubit.  The relaxation time of the nuclear magnetic moment of P in silicon was extensively studied in bulk resonance experiments in the 1960s\cite{Wilson_1961}, and is known to be in excess of 10 hours at low temperatures. This makes the system highly attractive with respect to decoherence issues.  Since the nuclei we wish to store our information on are so well isolated, read-out is extremely difficult.  

This difficulty in read-out can (in principle) be overcome using spin to charge transduction\cite{Kane_1998} as follows.  A P atom has five electrons in its outermost shell.  Four of these are used to bond to the surrounding Si lattice, leaving a single valence electron loosely bound to the P nucleus.  The spin orientation of this electron can be correlated with the nuclear spin through the hyperfine interaction.  In this way the electron can be used to address and infer information about the nuclear spin state.  The strength of the hyperfine interaction is determined by the electron's wavefunction at the nucleus.  By carefully controlled voltages on gates placed ontop of the Si substrate (A gates in Fig.\ref{fg:Kane_qc}), the electron/nucleus overlap can be controlled to bring the P nuclear magnetic moment into and out-of resonance with an external, global, oscillating magnetic field.  This allows for single qubit rotations.

\begin{figure}[b!]
\centering
\includegraphics[width=7cm]{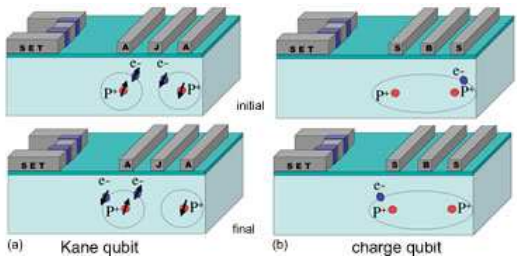}
\caption{Initial and final states of the read-out process for the Kane and charge-qubit SSQCs.}
\label{fg:qubit_read-out}
\noindent 
\end{figure}

The electron's wavefunction extends across many lattice sites.  By placing appropriate voltages on surface gates placed between qubits (J gates in Fig.\ref{fg:Kane_qc}), the wavefunction overlap of neighboring P dopants' electrons can, in principle, be controlled, allowing for controlled two qubit interactions.  In this way multi-qubit logic gates can be achieved.  In practice such control will require precise atomic placement of the P donors \cite{Koiller_2002,Wellard_2003}.  This use of the valence electron as a `messenger' between qubits can only occur if the decoherence time of the electron spin system is long enough to allow for information processing operations to occur.  At sufficiently low temperatures, electron dephasing times of greater than 60ms have been reported\cite{Tyryshkin_2003}, which is more than ample.  The Kane SSQC read-out (Fig.\ref{fg:qubit_read-out} (a)) and  allied electron-spin proposals\cite{Vrijen_2000,Hill_04,de_Soussa_2004} consists in transferring the valence electron of a P dopant to a neighboring P-dopant (forming a doubly occupied D$^-$ state on the latter site).  Originally this scheme involved transfer by a DC field.  To avoid ionization of the D$^-$ state a resonant transfer technique has been proposed\cite{Hollenberg_resonant_2004}.  This D$^-$ state can occur only if the two P donors have opposite nuclear spins.  This is due to carefully correlating the valence electron's spin to the nuclear spin by the hyperfine interaction, and the D$^-$ state being Pauli limited.  Spin to charge transduction is thus achieved, converting the difficult problem of single spin detection to charge detection.  In this way, by preparing a read-out qubit in a known spin orientation, the orientation of the qubit being measured can be inferred.  

The charge qubit SSQC (Fig.\ref{fg:charge_qc}) also uses P dopants in Si for qubits, however a P-$\rm P^{+}$ pair constitutes a single qubit.  In this way a single valence electron is shared between the dopants.  The \ket{0} and \ket{1} states are now encoded onto the location of the shared electron.  Read-out consists of determining this location (Fig.\ref{fg:qubit_read-out} b).  Geometries have also been proposed that may be sensitive to the superposition basis, by either adding an additional ionized donor \cite{Greentree_2004}, or placing the SET centrally between the donors\cite{Stace_2004}.  

Single qubit rotations for the charge qubit are achieved by varying voltages on control gates above the silicon substrate, in order to force the shared electron from the left (\ket{L}) to the right (\ket{R}) dopant nuclei.  Multi-qubit operations can be achieved by carefully placing the P-$\rm P^+$ pairs, such that the dynamics of one qubit becomes dependent on the other.  The silicon band-structure implies atomic precision placement of the P-P$^+$ pairs is required\cite{Hu_2004}. 

The disadvantage of the charge-qubit is that it is more susceptible to decoherence by local charge fluctuations due to the long range of the electromagnetic force.  This has been the subject of a number of theoretical investigations \cite{Barrett_2003,Fidichikin_2004}.  Current estimates put the decoherence rate in the order of $10\rm ns$ \cite{Barrett_2003}.  This short decoherence time will still allow for gate operations to occur due to the faster operation time (estimated to be of order 10 ps \cite{Hollenberg_2004}).

From the above description of both SSQC types it is apparent that the final read-out signal for both SSQCs is dependent on the location of a single electron on subsurface P donors.  Hence by using SETs to monitor the local electrostatic environment we can perform SSQC read-out.  Though our investigation is motivated by this specific application of SETs, the analysis technique we follow is general enough to be useful to any field intending to use the SET as an electrometer.  

\begin{figure}[tb!]
\centering
\includegraphics*[width=7cm]{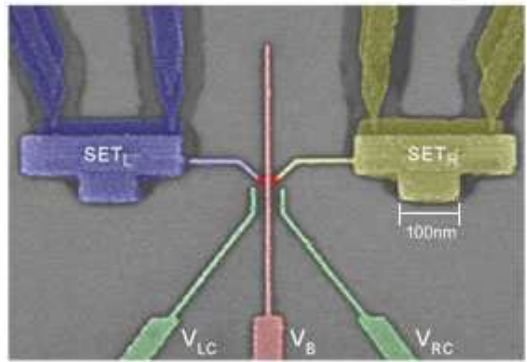}
\caption{SEM image (courtesy of F. Hudson et al, UNSW) of a typical experimental device providing motivation for the SET geometries under consideration.}
\label{fg:twinSET_STM}
\noindent 
\end{figure}

Our research is motivated by experiments in which single charge transfer between clusters of P-donors was monitored by two SETs\cite{Clark_2003,Buehler_2004} (Fig.\ref{fg:twinSET_STM}).  We use the structures of the experimental SETs as a starting point for our investigation into SET geometry optimization.  In keeping with the experimental architecture, our simulations consists of donors placed 20nm below a 5nm silicon oxide layer.  We also consider clusters of donors initially (see section \ref{exp_motive}), rather than single dopants, with the single dopant case considered in section \ref{cluster_to_single}. 

We begin our discussion by presenting some introductory SET theory, along with a description of the induced island charge as a measure of SET sensitivity.  The interplay between quantized electrons, and continuous electric fields in relation to understanding mesoscopic circuitry is then discussed.  The initial and final states of the read-out process in both types of SSQC is then described in detail.  In this paper we extend our previous description of SET sensitivity\cite{Pakes_2003} to calculate both the charge induced on a SET island, and the current variations that can be expected through the device due to charge transfer on nearby quantum dots (QD) (i.e. the nuclear or charge qubits).  We then  present a technique for calculating these currents in the steady state, based on a rate equation and geometric considerations of the SET.  The experimental setup which motivated our research, and the techniques we use to determine the capacitance matrix of our system are then presented.  Finally we present the results of simulations in which we varied the SET geometry in order to optimize its sensitivity as an electrometer.

\section{Single Electron Transistors}
\label{source}
\begin{figure}[tb!]
\centering
\includegraphics[width=7cm] {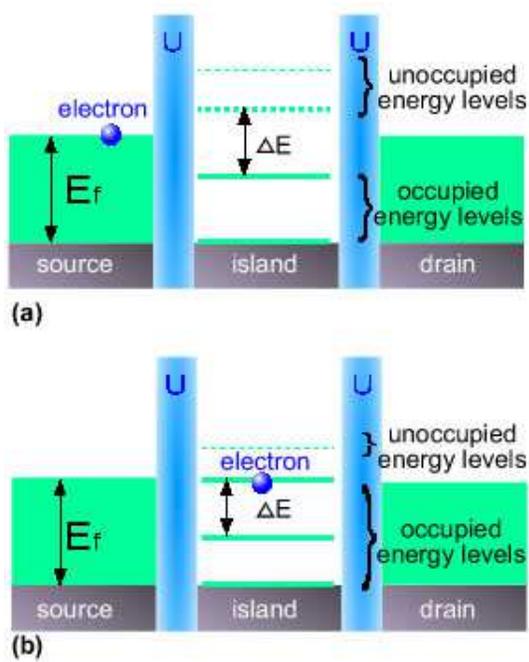}
\caption{ Discrete energy levels on a SET island.  For single electron tunneling to be observable, the energy spacing between levels must be greater than thermal fluctuations ($\Delta E>k_BT$).   (a) First unoccupied energy level on the island is greater than the Fermi energy ($E_f$) of the source preventing tunneling through the potential barrier U. (b) The energy levels of the island are shifted by the application of a nearby potential, allowing an electron to tunnel from the source.  The field effect of the excess electron on the dot produces a Coulomb blockade.  This prevents further electrons from tunneling onto to the island.}
\label{fg:Coulomb_block}
\noindent  
\end{figure}

As electronic devices approach the nanometer scale, the sea of continuous electronic energy levels normally present in metals becomes discrete and parts of the device begin to behave as low dimensional quantum systems.  When this occurs, the passage of single electrons through a tunnel barrier can be controlled.  By joining two tunnel barriers in series (effectively creating an isolated island), it is possible to create a transistor, the current though which is determined by single electron tunneling events.  The controlled tunneling of single electrons comes about due to the Coulomb blockade present on the SET island (Fig.\ref{fg:Coulomb_block}).  Any electron tunneling onto the island faces finding a particular energy level to inhabit, rather than the conduction band continuum normally occurring in metals. Tunneling phenomena of single electrons may be observed and controlled by varying the height of the island's energy levels through the application of a nearby electric potential (e.g. the gate of the transistor).  Thermal fluctuations in the device must be small enough not to override the resolution of the discrete energy levels, i.e. $\Delta E > k_BT$, where $\Delta E$ is the energy level spacing of the island.

The Coulombic repulsion of electrons present on the island prevents any further electrons from tunneling onto it.  Though the field produced by a single electron is quite small, the effect of this field on the device is inversely proportional to the size of the island.  Hence in the nanometer regime the field effects can become quite large.  This prevention of further electrons moving onto the island is referred to as the Coulomb blockade.  Only when an electron tunneling to the island has an energy equal to or greater than one of the unfilled discrete energy levels of the island is it able to overcome the Coulomb blockade.  By tuning the energy level on the island via the transistor gate, the energies for having $n$ or $n+1$ excess electrons on the island become degenerate.  Electrons will now able to tunnel sequentially from source to drain, creating a measurable current through the SET.

\begin{figure}[httb!]
\centering
\includegraphics[width=7cm]{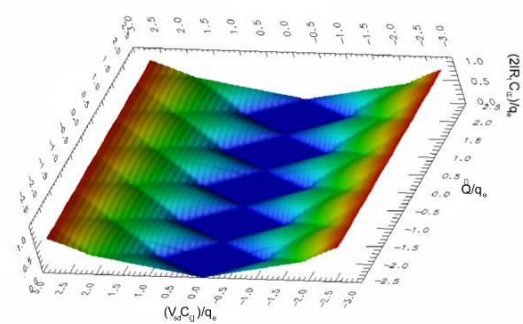}
\caption{Normalized current mapped across the gate-transport (source-drain) voltage plane (also normalized).  The characteristic Coulomb diamonds show stable regions of zero current.  Varying the gate voltage moves the SET island through successive states of excess electrons, while increasing the transport voltage reduces the Coulomb blockade, by introducing higher order charge states.}
\label{fg:Coulomb_diamonds}
\noindent 
\end{figure}

The behavior of the SET can be characterized by sweeping the gate voltage for a variety of source-drain (transport) voltages.  Mapping the current through the SET for each point in this gate-transport voltage plane produces regions of zero current in which the tunneling is suppressed, and the current is essentially zero.  These regions correspond to Fig.\ref{fg:Coulomb_block} (a), and have a characteristic diamond shape.  Each successive Coulomb diamond (in the gate voltage direction) represents a region of the parameter-space in which the charging energy of the system is minimized by adding a single electron to the island.  By increasing the transport voltage, higher charge states become available in the tunneling process, reducing the efficacy of the Coulomb blockade, as shown in Fig.\ref{fg:Coulomb_diamonds}.

\subsection{The SET as an Electrometer}
As described above, the variation of the potential applied to the SET gate varies the energy level on the island, changing the strength of the Coulomb blockade.  In an analogous fashion, any charge on a nearby QD will vary the island's potential.  We can associate a capacitive coupling between the island and the nearby QD.  The potential variation on the island is a continuous variable based on the island-QD capacitive coupling.  From the definition of capacitance ($C\equiv Q/V$) we can associate this potential variation with an induced charge $\delta q$, which may be a fraction of an electron charge.

The greater this $\delta q$, the stronger the variation in the energy levels on the island.  As such, this $\delta q$ is commonly used to describe the sensitivity of the SET to the system being measured.  The actual signal measured however is a current through the SET which we calculate in section \ref{SET_current}.

\subsection{The Induced Island Charge}
\label{induced_charge}
We present two methods of determining the charge induced on an island for the simplest system of a SET coupled to a single QD.  We extend the technique to a two QD system for our analysis, though present the single QD case here for clarity.  The motivation for developing two approaches is that the first is a common method in the literature, and is very simple, while the second allows us to determine the energy variations of the entire system due to tunneling events.  These energy variations can then be used to determine current variations in the SET.  Comparison of the two techniques indicated that both gave equivalent results for our simulations.

\begin{figure}[tb!]
\centering
\includegraphics[width=7cm]{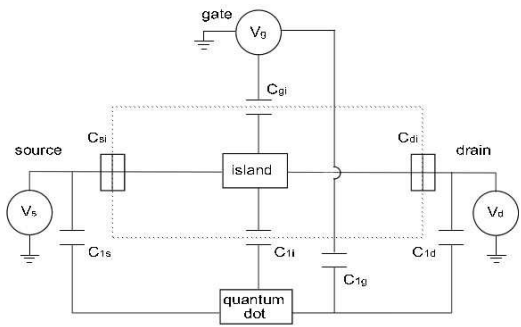}
\caption{Circuit diagram of the SET-QD system.  Isolated regions with discrete energy levels are represented by solid boxes.  The dotted outline indicates the region considered as the SET island.}
\label{fg:SETQD_circuit}
\noindent 
\end{figure}

The circuit diagram for the system being considered is shown in Fig.\ref{fg:SETQD_circuit}.

\subsubsection{Ratio of QD's Total Induced Charge}
\label{ratio_method}
Suppose we have a QD charged with $n$ excess electrons ($q=-nq_e$) coupled to a SET.  Let the self capacitance of the QD be $C_{1\Sigma}$.  This self capacitance is simply the summation of the QD's capacitive coupling to each of the objects in its environment.  In our model we assume the capacitive coupling to defects in the lattice (dangling bonds, local impurities etc.) is negligible so that we have,
\begin{equation}
\label{eq:dot_self_cap}
{C_{1\Sigma}}
=
C_{1i}+C_{1s}+C_{1d}+C_{1g}\;,
\end{equation}
where the roman subscripts refer to the SET's island, source, drain and gate.
Conservation of charge necessitates that the charge on the QD must induce a charge of $nq_e$ in its environment.  Hence we can write the manner in which this charge is shared across the environment as, \begin{equation}
\label{eq_charge_share}
nq_e=\left (
\frac{C_{1i}}{C_{1\Sigma}}
+
\frac{C_{1s}}{C_{1\Sigma}}
+
\frac{C_{1d}}{C_{1\Sigma}}
+
\frac{C_{1g}}{C_{1\Sigma}}
\right )nq_e\;.
\end{equation}
Inspection of Equation (\ref{eq_charge_share}) tells us immediately that the charge induced on the island due to the QD is simply
\begin{equation}
\delta q=\frac{C_{1i}}{C_{1\Sigma}}nq_e\;.
\end{equation}
 
\subsubsection{The Electrostatic Energy Variation of the System}
\label{electro_system}
This second method for determining $\delta q$, though more complicated than the previous, has the advantage of allowing for calculation of energy variations of the system due to tunneling events, which is essential for determining the current through a SET (see section \ref{SET_current}).

The electrostatic energy of a capacitor is given by  $E=q^2/(2C)$ where $q$ is the magnitude of the charge on one of the plates of the capacitor.  The electrostatic energy of the SET-QD system under investigation is simply the matrix equation extension of this, accounting for all the capacitances in the system. Any electrodes connected by wires to regions outside the system have their potentials controlled externally, and so we need only consider the charges on any floating electrodes (in this case the island and a single QD).  The charging energy for the system is given by
\begin{equation}
\label{eq:system_energy}
E=\frac{1}{2}Q^TC^{-1}_EQ\;.
\end{equation}
Q is a vector containing the charge on the island and the QD and we refer to $C_E$ as the `energy' capacitance matrix, defined below.  The charge on the island and QD is a combination of actual electrons, and induced charge due to voltage differences, 
\begin{equation}
Q=\tilde{Q}+\mathbf{ n}q_e\;,
\end{equation}
where \textbf{n} represents the number of electrons on each low dimensional quantum system (in this case the island and the subsurface QD), while $\tilde{Q}$ represents the continuous charge induced on these objects.  In matrix form we have,
\begin{equation}
Q=\left(\begin{array}{c}\tilde{Q_i} \\ \tilde{Q_1}\end{array}\right)+\left(\begin{array}{c}n_i \\n_1\end{array}\right)\;.
\end{equation}
Considering only the island for a moment, we can write the induced charge from the source, drain and gate electrodes as
\begin{equation}
\tilde{Q}_{i}=C_{is}V_s +C_{id}V_d + C_{ig}V_g\;,
\end{equation}
where we are referencing all voltages to the island ($V_s=-V_d$ assuming the tunnel barriers in our SET are identical).  
For our system then, we can write $\tilde{Q}=C_cV$.  We refer to $C_c$ as the `correlation' capacitance matrix,
\begin{equation}
C_c=
\left(\begin{array}{ccc}
C_{is} & C_{id} & C_{ig} 
\\C_{1s} & C_{1d} & C_{1g}
\end{array}\right) 
\;\;,\;V=\left(\begin{array}{c}V_{s} \\V_{d} \\V_{g}\end{array}\right)\;.
\end{equation}
where the subscript 1 again refers to the QD.  This can easily be extended to an arbitrary number of QDs.  We can now write the total charge on the QD and the island as a function of the electrode voltages and elements of the system's capacitance matrix, with the number of electrons on each as a free parameter we will vary in order to find the minimum charging energy of the system.

The $C_{E}$ matrix describes the cross capacitances of the QD and the island and is given by,
\begin{equation}
C_{E}=\left(\begin{array}{cc}
C_{i\Sigma} & -C_{i1} 
\\-C_{1i} & C_{1\Sigma}
\end{array}\right)\;,
\end{equation}
where C$_{\i\Sigma}$ is defined in Equation (\ref{eq:dot_self_cap}).  The island's self capacitance (C$_{i\Sigma}$) has an equivalent form. 

While not necessary to the theory, it is worth noting that this matrix must be symmetric, which can reduce computation time for more complicated systems.  We are now in a position to calculate the electrostatic energy of the system for arbitrary charge configurations on the QD and island by simply varying the integers constituting \textbf{n}.  From this the $\delta q$ induced on the island by the presence of a charge on the QD will be determined.  We do this for the case of a single electron being added to the QD, and determine $\delta q$ by the shift in the system's electrostatic energy profile.

\begin{figure}[tb!]
\centering
\includegraphics[width=7cm]{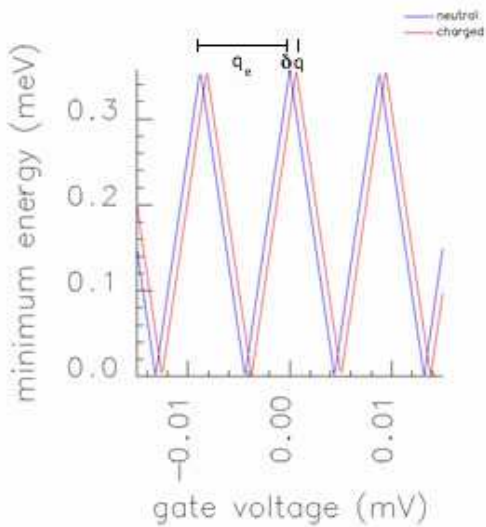}
\caption{Plot of difference between two lowest energy levels on the island.  The blue line is for the quantum dot being measured neutral, the red line for a charge of $-q_e$ on the quantum dot.  Peak to peak distance corresponds to a charge of $q_e$, enabling $\delta q$ to be determined by the shift between the red and blue plots.}
\label{fg:dE_plot}
\noindent 
\end{figure}

To determine the shift in the system's electrostatic energy profile we first determine the two lowest energy levels of the system for the case of the QD being neutral.  We set the source and drain voltages to $+V_{sd}/2$ and $-V_{sd}/2$ (in our simulations $V_{sd}$ = 0.1 mV) and step through a voltage range on the bias gate $\pm V_{gmax}$.  For each applied bias voltage ($V_g$) we cycle through $\pm n$ excess electrons on the island  and use Equation (\ref{eq:system_energy}) to determine the electrostatic energy of the system.  We then keep the difference between the two lowest energies and plot this energy difference with respect to the bias gate's voltage (see Fig.\ref{fg:dE_plot}).

The zeros of this plot are voltages at which the island having $n$, or $n+1$ excess electrons is degenerate with respect to the energy of the system, and correspond to maximum currents through the SET.  Similarly, the period of the plot corresponds to adding one electron to the island.

To determine $\delta q$ we then perform the entire process again, this time placing one electron on the QD.  This will give a similar energy difference plot, but shifted due to the dot's influence on the island.  Since we know the period of the plot corresponds to an electron being added to the island, we can thus deduce what fraction of an electron corresponds to the shift, giving $\delta q$.

\section{Current Through a SET}
\label{SET_current}
Though we have focused on the induced island charge ($\delta q$) as characterizing the sensitivity of our SET when measuring our QD system, the actual signal detected in experiments is a variation of current.  Having demonstrated how to determine the charging energy of our system in section \ref{electro_system}, we now show how to extend the method to determine actual currents through the device.  Our technique closely follows Grabert and Devoret\cite{Grabert_1992} and gives currents in the nA regime, which is in agreement with the measured results for SETs in general\cite{Fulton_1987,Grupp_2001}.

To determine the current through the SET we need to consider the tunneling rates on and off the island.  The rate is determined by the energy difference of the system between different configurations of electrons on the island.  We make the assumptions that  co-tunneling events can be ignored, and assume that only the island has quantised energy levels.

The rate at which electrons tunnel from the electrodes to the island is dependent on the number of excess electrons on the island.  Given the island is in a state with $n$ excess electrons, and at a temperature $T$ we denote the rate at which electrons tunnel from the island ($i$) to the electrode $\chi$ (either the source or drain) as, 

\begin{equation}
\label{eq:rate}
\Gamma_{\chi i}^n=
\frac{1}{q_{e}^{2}R_{t}}
\frac{\Delta E_{\chi i}^n}{\exp\left (\frac{\Delta E_{\chi i}^n}{k_{B}T} \right )-1}\;.
\end{equation}

We used 4K for our simulations to avoid low temperature convergence issues in determining the capacitive coupling between the objects.  In practice the operational temperature of a SSQC would be of order mK, though this won't effect the qualitative results of our analysis.  The parameters of interest are the change in energy ($\Delta E_{\chi i}^n$) and the barrier tunnel resistance ($R_t$).  

We determine $\Delta E_{\chi i}^n$ by using Equation (\ref{eq:system_energy}) for the charging energy of the system, and considering the work done (either on or by the system) in pushing an electron through the tunnel barrier.  The energy variation for Equation (\ref{eq:rate}) is given by:
\begin{equation}
\label{eq:deltaE}
\Delta E^{n}_{\chi i} = E(n-1) - E(n) + V_\chi q_e\;,
\end{equation}  

For the tunnel resitance we use the expression\cite{Schon_1997}
\begin{equation}
\label{eq:resitance}
R_{t}=
\frac{\hbar^3\exp(\frac{2W}{\hbar}\sqrt{2m_e\phi})}
{2\pi m_e^* q_{e}^2 E_F A}\;,
\end{equation}
where W is the width of the tunnel barrier and A is its surface area.  The variable $\phi$ is the height of the potential barrier (taken to be a typical value of 2eV\cite{Koppinen_2003} and $E_F$ is the Fermi energy of the island (taken to be 11.65eV at 4K\cite{Brenner_2004}).  The effective mass of an electron in aluminium oxide is taken to be $m_e^*=0.35m_e$\cite{Xu_1991}. 

To obtain the current through the SET we must consider the contribution from all possible processes.  Since we are ignoring  co-tunneling processes we need only consider events that change the number of electrons on the island by 1. The current is then given by
\begin{equation}
\label{eq:current}
I=q_{e}\sum_{n=-\infty}^{\infty}\sum_{\chi = s,d}p_{n}(\Gamma^{n}_{\chi i} -\Gamma^{n}_{i \chi})\;,
\end{equation}
where $p_n$ is the probability that the island is in a state with $n$ excess electrons.  To determine these probabilities we consider the master equation
\begin{eqnarray}
\label{eq:master}
\dot{p}_n & = & (\Gamma^{n-1}_{is}+\Gamma^{n-1}_{id})p_{n-1}
\\ & &\nonumber  \mbox{}-(\Gamma^n_{is}+\Gamma^n_{si}+
\Gamma^n_{di}+\Gamma^n_{id})p_n
\\ & &\nonumber  \mbox{}+(\Gamma^{n+1}_{si}+\Gamma^{n+1}_{di})p_{n+1}\;.
\end{eqnarray}
Equation (\ref{eq:master}) simply states that the rate of change of occupation number must simply be the rate at which electrons are entering the state, less the rate at which they are leaving.  

\begin{figure}[tb!]
\centering
\includegraphics*[width=7cm]{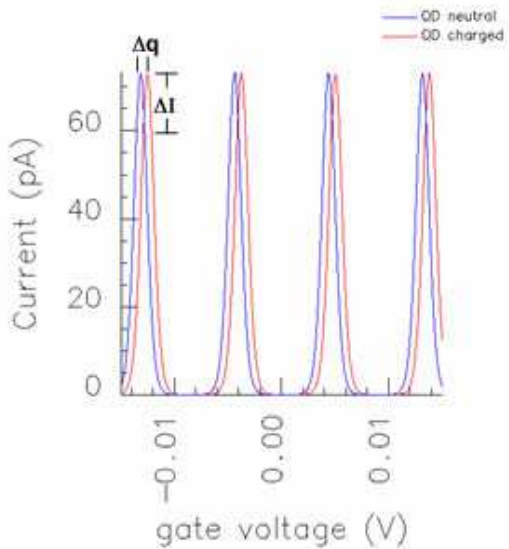}
\caption{Example of currents calculated using the above method.  Blue plot is for measured QD being neutral, red is for the QD having $-q_e$ excess charge.  The charge on the QD induces a shift in the current ($\delta I$), due to the charge induced on the island by the QD ($\delta q$) behaving in an analagous fashion to a voltage variation on the gate.}
\label{fg:current_example}
\noindent 
\end{figure}

In order to solve the problem numerically we must truncate the summation.  We let the island charge configuration range over $\pm N$ (with $N=10$), and Equation (\ref{eq:master}) becomes a system of $2N+1$ equations.  We found $N=10$ to be sufficient for convergence, with $N=5$, $N=10$ and $N=20$ all giving the same results for Equation (\ref{eq:current}).  We write these $2N+1$ coupled equations in a single matrix equation.  For clarity we make the substitutions
\begin{equation}
\nonumber
\Gamma^{n-1}_{is}+\Gamma^{n-1}_{id} = A_n,
\end{equation}
\begin{equation}
\Gamma^n_{is}+\Gamma^n_{si}+
\Gamma^n_{di}+\Gamma^n_{id} = B_n,
\end{equation}
\begin{equation}
\nonumber
\Gamma^{n+1}_{si}+\Gamma^{n+1}_{di} = C_n\;.
\end{equation}
Notice that $A_{-N}$ contains the probability for the $-N-1$ state, and $C_N$ for the $N+1$ state.  We must therefore make the further approximation that $A_{-N} = 0$ and $C_N = 0$.  The problem can be simplified further by recognizing that we are after the current in the steady state, and the $p_n$ are probabilities, hence
\begin{equation}
\dot{p}_n=0\;,
\end{equation}
and
\begin{equation}
\label{eq:normalization}
\sum_n{p_n}=1\;.
\end{equation}
The matrix equation can therefore be written as (replacing the final equation with the normalization condition of Equation (\ref{eq:normalization}))
\begin{widetext}
\begin{equation}
\label{eq_rate_matrix}
\left(\begin{array}{c}
 0\\ 0\\ 0\\ 0\\ 0\\ 0\\ 1
\end{array}\right)=
\left(\begin{array}{ccccccc}
-B_{_{-N}} & C_{_{-N}} & 0 & \cdots & \cdots & \cdots & 0 
\\ A_{_{-N+1}} & -B_{_{-N+1}} & C_{_{-N+1}} & \cdots & \cdots  & \cdots & 0 
\\\vdots & \vdots & \vdots & \vdots & \vdots & \vdots & \vdots 
\\ 0 & \cdots & A_{n} & -B_{n} & C_{n} & \cdots & 0 
\\\vdots & \vdots & \vdots & \vdots & \vdots & \vdots & \vdots 
\\0 & \cdots & \cdots & \cdots & A{_{_N-1}} & -B_{_{N-1}}& C_{_{N-1}}
\\1 & 1 & \cdots & \cdots & \cdots & 1 & 1 
\end{array}\right)
\left(\begin{array}{c}
p_{_{-N}}
\\p_{_{-N+1}}
\\\vdots
\\p_n
\\\vdots 
\\p_{_{N-1}}
\\p_{_{N}}
\end{array}\right)\;.
\end{equation}
\end{widetext}
Equation (\ref{eq_rate_matrix}) is a simple matrix equation of the form X=YP, which we solve for P, giving all the occupation probabilities.  Inserting these probabilities into Equation (\ref{eq:current}) we thus obtain the current through the SET.  Fig.\ref{fg:current_example} shows an example of the currents calculated using this method.  It clearly displays the expected Coulomb blockade regions.  The two plots are the currents for the QD neutral, and having charge of $-q_e$.  The shift in the plots is due to the charge induced on the island due to the QD charge.  When performing measurements of a SSQC, the signal will be the variation in the current, due to the shift brought about by the charge on the quantum dot.  Our results below are thus displayed as the difference between the neutral and charged plots.  We can maximize our signal by making measurements when the SET current is at a maximium rate of change with respect to voltage variation on the gate (i.e. at maximum transconductance).

\section{Determining Capacitances}
\subsection{Experimental Motivation}
\label{exp_motive}
\begin{figure}[tb!]
\centering
\includegraphics*[width=7cm]{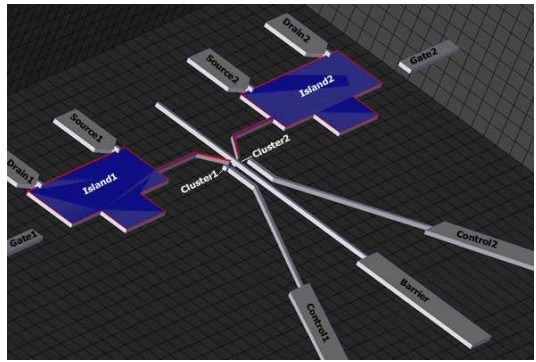}
\caption{ Geometry of experimental motivation for SET geometries under investigation.  Control and barrier electrodes induce charge motion between the clusters.  SET gates compensate the effect of the control and barrier electrodes on the SETs' currents, allowing the detection of charge motion between the clusters.}
\label{fg:UNSW_geo}
\noindent 
\end{figure}
\begin{figure}[tb!]
\centering
\includegraphics[width=7cm]{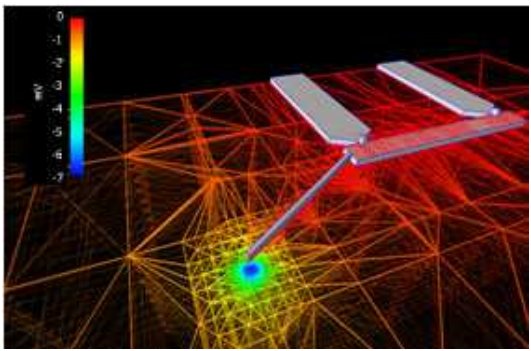}
\caption{Example of mesh variation based on geometry.  The colours represent the electrostatic potential due to an excess electron placed on a subsurface quantum dot.  Finer meshes near metallic objects are generated automatically to capture the rapidly spatially varying electric field.}
\label{fg:mesh_example}
\noindent 
 
\end{figure}

We base our SET geometry investigation around recent experiments\cite{Clark_2003}.  The experiment involves two Al SETs (with oxide tunnel barriers) atop a 5nm $\rm SiO_2$ layer deposited on high resistivity Si.  The SETs are coupled to a double QD system embedded in the Si substrate.  Control and barrier gates are used to mediate the electron transfer process between the QDs.  The gates of the SETs compensate the effect of the control and barrier electrodes on the SETs' islands (See Fig.\ref{fg:UNSW_geo}).  This ensures that only uncompensated effects due to electron motion between the QDs being measured give a variation in the currents through the SETs.  Once the full capacitance matrix of the system is determined, we use this to calculate read-out signals using the above arguments.  To determine the capacitance matrix of such a complicated geometry, it is necessary to use the numerical technique of finite element modeling (FEM).  In FEM, the physics of a system being investigated is calculated on a mesh which discretises the geometry of the situation.  To reduce computing requirements careful consideration needs to be given the mesh construction.  Finer meshes are required in regions where the electric field is expected to vary rapidly.  Fig.\ref{fg:mesh_example} shows an example of this for the single SET single QD system discussed in section \ref{electro_system}.  We use ISE-MESH\cite{TCAD} and in-house computational geometry algorithms to generate the mesh automatically based on our input geometery.  The experiments motivating our research employed twin clusters of ~600P donors as a test for single electron transfer.  We thus modeled our QDs as metalic cubes ($60\times 60\times60 \rm nm^{3}$) with a surface area of approximately the same size as the spheroids determined by SRIM\footnote{SRIM: Stopping and Range of Ions in Matter computer model based on J.F.Ziegler, J.P. Biersack and U. Littmark - The Stopping and Range of Ions in Solids, Pergamon Press, New York, (1985).} calculations of the ion implantation process\cite{Jamieson_2003} used to make the clusters.  In section \ref{cluster_to_single} we investigate the change in the measured signal through the SET due to measuring single donors rather than clusters. 

\subsection{Capacitance Calculation}
Once the mesh is determined, capacitances are determined by performing an AC analysis through the device using ISE-DESSIS\cite{TCAD}.  This constitutes placing a known alternating current between the objects we wish to know the capacitance of and solving
\begin{equation}
\label{eq:currentAC}
I=A\cdot V+j\cdot \omega \cdot C \cdot V\;,
\end{equation}
$I$  is  the vector containing the current at each of the nodes (determined by applying Kirchhoff's current law), $V$  is  the vector containing the known potential at each node, $A$  is the admittance matrix containing the resistance between each node, $\omega$ is the frequency of the AC.  We use a frequency of 1MHz, but the model is not sensitive to this parameter since we have a pure Si substrate. Finally $C$ is the capacitance matrix containing the capacitances in the device.  It is worth noting that some properties of the capacitance matrix can serve as a check for the correct behavior of the simulation.  Firstly the elements of the capacitance matrix ($C_{\alpha \beta}$) denote the capacitance between the $\alpha$th and $\beta$th objects hence $C_{\alpha \beta}=C_{\beta \alpha}$.  Diagonal terms ($C_{\alpha \alpha}$) denote the total self capacitance of the $\alpha$th object.  Since the charge on the $\alpha$th object induces  a charge of opposite sign on all other objects:
\begin{equation}
-\sum_{\alpha \neq \beta}{C_{\alpha \beta}}=C_{\alpha \alpha}\;.
\end{equation}

\section{Geometries Investigated}
\begin{figure}[tb!]
\centering
\includegraphics[width=7cm]{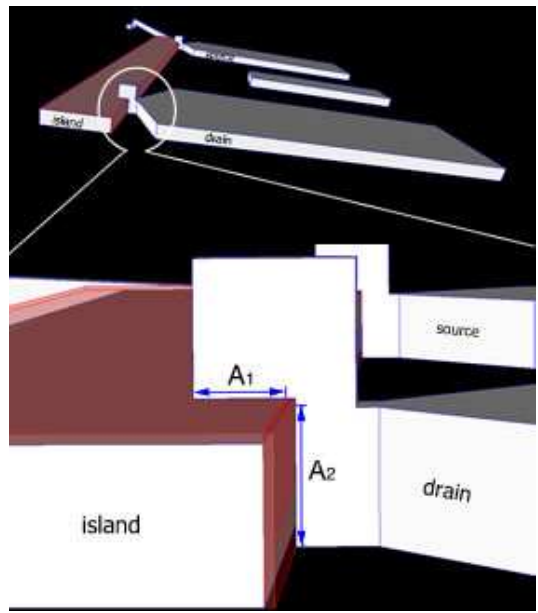}
\caption{Geometry of tunnel barrier.  The vertical and horizontal areas contribution to the SET current calculated separately and added.}
\label{fg:barrier_geo}
\noindent 
\end{figure}
The effects of misalignment of the QDs to the twinSET architecture has been considered in previous studies\cite{Lee_2004} using the FASTCAP package, though the geometry of the SETs was not investigated.  Due to the computational requirements of FEM, we were not able to perform simulations of the entire twinSET device that motivated our investigations (see section \ref{exp_motive}).  We instead chose to model a single SET based on the experimental set-up, coupled to two QDs.  While this does not change the qualitative analysis of our results, we do expect the signals for the full twinSET system to be smaller due to the larger number of metallic elements the charge we wish to measure can couple to.  When modeling the Kane qubit the SET is coupled to two QDs.  Initially the current are determined with both QDs neutral.  The final configuration is then with one QD having $q_e$ and the other having $-q_e$.  For a charge qubit the SET is also coupled to a two QDs, one with the charge of a hole, which changes its location (see Fig.\ref{fg:qubit_read-out}).

For both qubit types we investigated four aspects of the SET geometry.  Firstly, we grew the antenna from the island towards the QD to confirm that this improved the signal for the experimental device.  We then repeated this setup with the gate placed between the island and the drain, in order to shield the island's antenna from the gate.  An investigation into the signal variation for increasing the size of the island was then performed, as this will vary its capacitive coupling to the rest of the environment.  Finally we investigated varying the overlap of the source and drain with the island as this varies the tunnel barrier area, implying an increase in the detected current variation.  The geometry variation can be seen in Fig.\ref{fg:antenna_geo}, Fig.\ref{fg:island_grow}, and Fig.\ref{fg:barrier_grow}.  For the geometries investigated the barrier wraps over two faces of the island.  We treat this by calculating the resistance, width and area of the vertical and horizontal parts individually, adding the results for the current (see Fig.\ref{fg:barrier_geo}).  

\subsection{Results of Growing Antenna}
\begin{figure}[tb!]
\centering
\includegraphics[width=7cm]{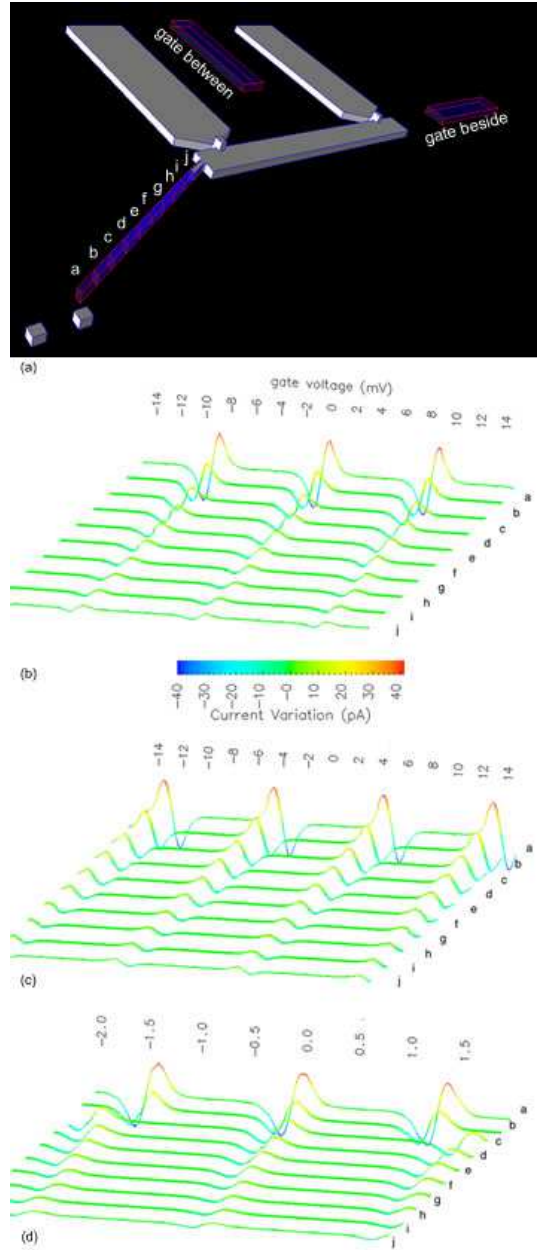}
\caption{(a) Geometries for a series of simulations investigating the effect of an antenna on the SET island.  Each increment of the antenna length is 10nm.  Simulations were run with the gate beside the drain, and then between the source and drain.  (b) Current variation for growing antenna with SET gate beside drain when measuring a charge qubit read-out event. (c) Current variation for growing antenna with gate beside drain when measuring a kane qubit read-out event. (d) Current variation for growing antenna with gate between the source and drain when measuring a charge qubit read-out event.}
\label{fg:antenna_geo}
\noindent 
 
\end{figure}

Fig.\ref{fg:antenna_geo}(a) shows the series of geometries investigated for growing an antenna from the island to the two QDs representing the measured qubit.  Increments of the antenna length for each simulation was 10nm.  The procedure was carried out for the gate placed beside the drain, and for the gate placed between the source and drain.

Fig.\ref{fg:antenna_geo}(b) and (c) show the variation in current through the SET with respect to the gate voltage.  The variation is for the initial and final qubit read-out states for the charge and Kane qubits respectively, for each antenna length shown in Fig.\ref{fg:antenna_geo}(a).  Note that for the purpose of comparison we have kept the spacing between QDs constant for all simulations, although donor spacing for the Kane and charge SSQC will most likely be different.  The results clearly show an improvement in the detected signal for the read-out event as the antenna is extended towards the system being measured.  The signals for both the charge and Kane qubit read-out events are of the same size, with the different qubit types simply giving a shift in the required gate voltage for maximizing the detected current variation.   This trend was consistent for all the simulations performed and henceforth only results for the charge qubit will be discussed.

Fig.\ref{fg:antenna_geo} (d) shows the effect of placing the gate between the source and drain.  The size of the current variation is the same for both cases (charge qubit data displayed in figure), but the periodicity is greatly increased for the between case (note the change in the voltage scale between Fig.\ref{fg:antenna_geo} (c) and Fig.\ref{fg:antenna_geo} (d)).  This is due to the self capacitance of the island increasing when the gate is placed between the source and drain.  Though the signal size is not affected, the increase in periodicity is not desirable as it would make the current in the SET more susceptible to voltage fluctuations in the gate.

\subsection{Results of Increasing Island Size}
\begin{figure}[tb!]
\centering
\includegraphics[width=7cm]{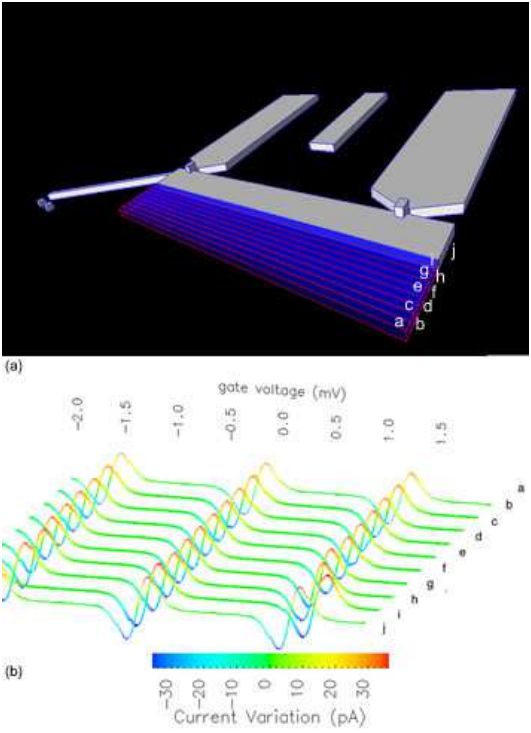}
\caption{(a) Geometries for a series of simulations investigating the effect of the SET island size.  Each simulation increased the width of the island by 5nm.  (b) Current variation for increasing island size when measuring a charge qubit read-out event.}
\label{fg:delta_I_island_charge}
\label{fg:island_grow}
\noindent 
 
\end{figure}

Fig.\ref{fg:island_grow}(a) shows the geometries used in a series of simulations investigating the effect of increasing the size of the island.  The width of the island was increased by 5nm for each simulation.  Though the width of the island was more than doubled, no significant change in the current variation of the SET due to a read-out event was shown (Fig.\ref{fg:delta_I_island_charge}(b)).  A slight increase in the periodicity of the signal did occur.  Of course the island must be small enough so that the energy level separation due to charging events are greater than the average thermal energy of the electrons in the system.  Our results show that this occurs for a wide range of island sizes.  As such the size of the island is not a significant parameter for consideration in SET design.  It should be noted however that our simulations did not include any stray capacitances to other objects in the environment near the SET.  In reality we expect random impurities in the silicon and silicon oxide regions, to which the island will capacitively couple, reducing the signal size, implying a smaller island size may be desirable.

\subsection{Results of Varying Tunnel Barrier}
\begin{figure}[b!]
\centering
\includegraphics[width=7cm]{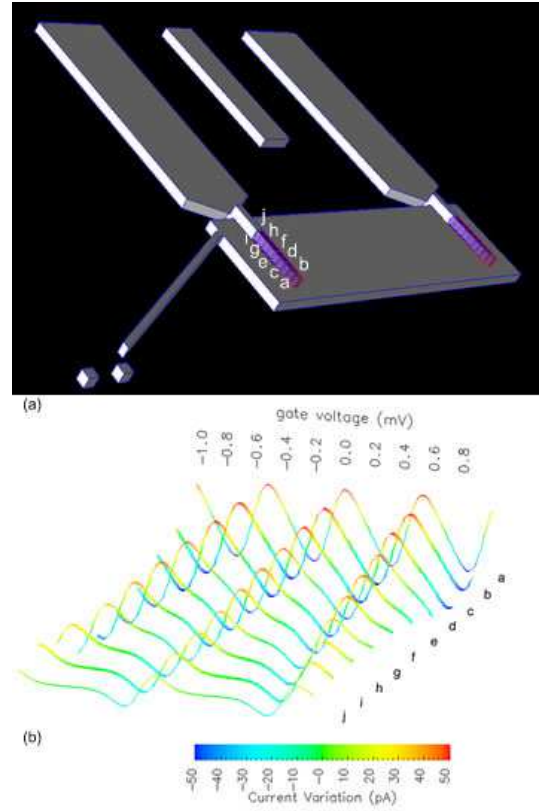}
\caption{(a) Geometries for investigating the effect of increasing the overlap of the source and drain with the island. The overlap was increased by 5nm for each simulation.  (b) Current variation for increasing overlap when measuring a charge-qubit read-out event.}
\label{fg:barrier_grow}
\noindent 
\end{figure}

\begin{figure}[b!]
\centering
\includegraphics[width=7cm]{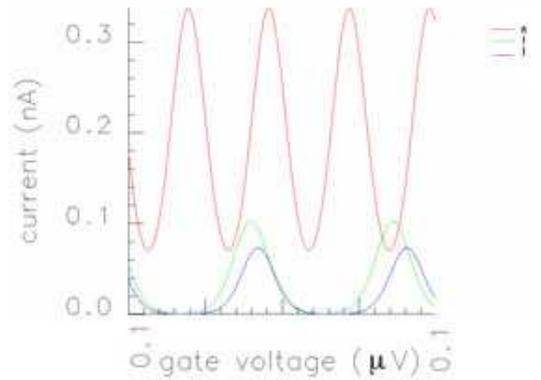}
\caption{Currents through SET for geometries a,i and j in Fig.\ref{fg:barrier_grow}.  Increasing the barrier area suppresses the Coulomb blockade.}
\label{fg:I_barrier}
\noindent 
 
\end{figure}

Fig.\ref{fg:barrier_grow} (a) shows the geometry for a series of simulations investigating the effect of increasing the overlap of the source and drain with the island.   The overlap was increased by 5nm for each simulation.  We expect to see an increase in the current through the SET as the overlap is increased, as more area is available for electrons to tunnel through.  This is clearly shown in Fig.\ref{fg:barrier_grow} (b) which displays the current variation for a charge qubit read-out event for the geometries in Fig.\ref{fg:barrier_grow} (a).  Although an increase in current is desirable as it gives a larger detectable signal, note that for large overlaps, the Coulomb blockade no longer completely suppresses  the current through the SET.  This is shown in Fig.\ref{fg:I_barrier} where the larger tunnel area simulation always has some current passing though it.  This would mean that during all qubit operations the SET would have a non-zero current.  This would be detrimental to the operation of the SSQC, as the back-action of the SET current on information processing qubits would contribute to the overall decoherence.  We found that simulation `i' in Fig.\ref{fg:barrier_grow} was the maximum overlap (~30nm $\times$ 10nm +10nm $\times$ 10nm) that could be used before this occurred.

\subsection{Results of using Single Donors instead of Clusters}
\label{cluster_to_single}
The previous analysis has been based on measuring QDs based on clusters of $\sim{\!600}$ P donors.  Based on data from SRIM calculations the QDs were modeled as 60$\times$60$\times$60 $\rm nm^3$ metallic cubes.  In order to model the single donor case the previous simulations were repeated with the original QDs replaced with 6$\times$6$\times$6 $\rm nm^3$ QDs.  6nm was chosen based on the Bohr radius of a P donor valence electron in Si being $\sim{\!3}$nm.

\begin{figure}[tb!]
\centering
\includegraphics[width=7cm]{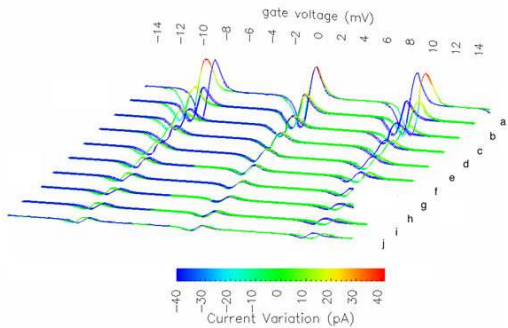}
\caption{Current variation for charge qubit read-out event with SET gate beside SET drain (as in geometry displayed in Fig.\ref{fg:antenna_geo} (a)).  The blue lines are the same as Fig.\ref{fg:antenna_geo} (b) with the coloured lines for the same simulation with the quantum dots being measured being replaced with 6$\times$6$\times$6 $\rm nm^3$ structures instead of 60$\times$60$\times$60 $\rm nm^3$ structures.  The plots indicate that the signal will be of the same size when performing measurements with single donors rather than clusters of $\sim{\!600}$.}
\label{fg:delta_I_cluster_single}
\noindent 
\end{figure}

The results of this are displayed in Fig.\ref{fg:delta_I_cluster_single} and show that the reduction in QD size has not changed the size of the detected signal, merely the required voltage for maximizing the signal.  This initially seems contradictory to the data in Lee et al.\cite{Lee_2004}, which found a variation in the induced charge for larger QD.  This work differs in that the variation in the current due to a transfer event, rather than simply the charge induced on an island due to a two QDs is taken as the cause of the variation in the detected signal.  Both QDs can capacitively couple more strongly to the island\cite{Lee_2004}, while the charge variation on the island due to an electrons motion between them remains unchanged.

\section{Conclusion}
We have investigated the effect of SET geometry variations on the detected signal for measurement of a double quantum dot system.  The two QD system represents the read-out process for two types of SSQC qubits, the charge and spin qubits.  The two qubit types are differentiated by the initial and final charge configurations on the QDs (see Fig.\ref{fg:qubit_read-out}).  QDs of size 60$\times$60$\times$60 $\rm nm^3$ were used initially, based on experiments using clusters of P donors.  We found that growing an antenna to couple the island to the QDs being measured greatly increased the detected signal for qubit read-out events of both types.  Variation in qubit types caused no change to the size of the signal, and only caused a shift in the required gate voltage for maximum current variation.  This was found to be the case for all geometry variations studied.

The position of the gate was moved from beside the island to between the source and drain.  This resulted in an increase in the periodicity of the current fluctuations through the SET.  Since this resulted in no increase in the actual signal size, we conclude that placing the gate to the side would be more desirable, as the SET would be more stable with respect to voltage fluctuations on the gate.

The size of the island was then increased to more than twice its original width.  This also increased the periodicity of the current fluctuations, although the effect was negligible.  Hence, (ignoring stray capacitances to local impurities) the size of the island is not a sensitive parameter in this architecture, as long as the energy spacing of the levels on the island remain greater than the thermal energy of electrons tunneling through the SET.

We then varied the overlap of the source and drain with the island, effectively increasing the area through which electrons can tunnel on and off the island.  Increasing this overlap increases the signal of read-out events, however if the overlap becomes to large, the Coulomb blockade no longer completely suppresses currents through the SET.  To avoid excess decoherence from the SET when measurements are not required, the SET will need to be switched on and off.  As such, the size of the overlap needs to be as large as possible while still allowing for the SET to be reliably turned off.  For widths of 10nm we found that the maximum overlap allowable was 30nm.

Finally we repeated all the simulations with QD of 6$\times$6$\times$6  $\rm nm^3$ instead of 60$\times$60$\times$60 $\rm nm^3$ in order to investigate the affect of using single donor atoms instead of clusters of $\sim{\!600}$.  The size of the signal we expect to see was not reduced at all, with the only effect being a shift in the required gate voltage for maximum signal detection.

We wish to thank K. Lee (UNSW) for helpful discussions with respect to experimental SET geometries.  We also wish to thank F. Hudson and the measurement team at UNSW for the SEM image of the twinSET device.  This work was supported by the Australian Research Council, the Australian government, the US National Security Agency, the Advanced Research and Development Activity and the US Army Research Office under contract number DAAD19-01-1-0653.

\bibliography{SET_sensitivity}

 \end{document}